\documentclass[aps,floatfix,prd,twocolumn,a4paper,10pt,citeautoscript,
longbibliography,showpacs]{revtex4-1}
\usepackage[english]{babel} 	
\usepackage{amsmath}        	
\usepackage{amssymb}        	
\usepackage{bm}				
\usepackage{bbm}            	
\usepackage{empheq}

\usepackage{float}       	
\usepackage{graphicx}       	
\usepackage{graphics}       	
\DeclareGraphicsExtensions{.jpg} 
\usepackage{hyperref}
\hypersetup{colorlinks=true, pdfstartview=FitV,linkcolor=blue, citecolor=blue, urlcolor=blue}

\newcommand{\Equation}[2]{\begin{equation}\label{#1}#2\end{equation}}
\newcommand{\Align}[2]{\begin{align}\label{#1}#2\end{align}}
\newcommand{\SubAlign}[2]{\begin{subequations}\label{#1}\begin{align}#2\end{align}\end{subequations}}

\newcommand{\bs}{\boldsymbol}
\newcommand{\Figref}[1]{Fig.~\ref{#1}}
\newcommand{\Eqref}[1]{\eqref{#1}}
\newcommand{\groupU}[1]{\mathrm{U}(#1)}  
\newcommand{\groupZ}[1]{\mathbb{Z}_{#1}} 
\newcommand{\groupO}[1]{{O}(#1)}         
\newcommand{\groupCP}[1]{\mathrm{C}\mathbb{P}^{#1}} 
\newcommand{\Exp}[1]{\text{e}^{#1}}

\renewcommand\Im{\mathrm{Im}}
\newcommand{\Grad}{{\bs\nabla}}

\newcommand{\Curl}{{\bs\nabla}\times}

\newcommand{\A}{{\bs A}}
\newcommand{\B}{{\bs B}}
\newcommand{\D}{{\bs \Pi}}

\newcommand{\F}{\mathcal{F}}
\newcommand{\Q}{{\cal Q}}
\newcommand{\J}{{\bs J}}

\newcommand{\Hc}[1]{\mathrm{H}_{c#1}}

\newcommand{\bvarphi}{{\bar\varphi}}

\begin{document}
\title{Microscopically derived multi-component Ginzburg-Landau theories for 
\texorpdfstring{$s+is$}{sis} superconducting state}
\author{Julien~Garaud}
\affiliation{Department of Theoretical Physics and  Center for Quantum Materials,
KTH-Royal Institute of Technology, Stockholm, SE-10691 Sweden}
\author{Mihail~Silaev}
\affiliation{Department of Theoretical Physics and  Center for Quantum Materials,
KTH-Royal Institute of Technology, Stockholm, SE-10691 Sweden}
\author{Egor~Babaev}
\affiliation{Department of Theoretical Physics and  Center for Quantum Materials,
KTH-Royal Institute of Technology, Stockholm, SE-10691 Sweden}

\begin{abstract}

Starting with the generic Ginzburg-Landau expansion from a microscopic $N$-band 
model, we focus on the case of a 3-band model which was suggested to be relevant 
to describe some iron-based superconductors. This can lead to the so-called $s+is$ 
superconducting state that breaks time-reversal symmetry due to the competition 
between different pairing channels. Of particular interest in that context, is 
the case of an interband dominated pairing with repulsion between different bands. 
For that case we consider in detail the relevant reduced two-component Ginzburg-Landau 
theory. We provide detailed analysis of the ground state, length scales and topological 
properties of that model. 
\newline{\it Prepared for the proceedings of Vortex IX conference in Rhodes (Sept. 2015).}

\vspace{1.5cm}

\end{abstract}

\date{\today}
\maketitle

\section{Introduction}

In many superconductors, the pairing of electrons is supposed to take 
place in several sheets of a Fermi surface which is formed by the overlapping 
electronic bands. To name a few, this is for example the case of MgB$_2$ 
\cite{MultibandMgB2}, Sr$_2$RuO$_4$ \cite{SrRuO,Kamihara.Watanabe.ea:08} 
or in more recently discovered iron-based superconductors \cite{Pnictides2,
Pnictides3,Pnictides4}.
Properties of superconductors in multiband systems can be qualitatively 
different from their simplest single-band $s$-wave counterparts. 
Of particular interest are the states that break additional symmetries.
Such states can appear if the superconducting gap functions phase differences 
between the bands differ from $0$ or $\pi$ \cite{Balatsky:00,Lee.Zhang.Wu:09,
Zhang2,StanevTesanovic,Fernandes,Agterberg,Ng,Lin,Johan,Bobkov,Chubukov2,
ChubukovMaitiSigrist}. Indeed in addition to the breakdown of the usual 
$\groupU{1}$ gauge symmetry, such superconducting states are characterized 
by an additional broken (discrete) time-reversal symmetry (BTRS). 
Spontaneous breakdown of the time-reversal symmetry has various interesting 
physical consequences, many of which are currently being explored. Iron-based 
superconductors \cite{Kamihara.Watanabe.ea:08} are among the most promising 
candidates for the observation of time-reversal symmetry breaking states 
that originate in the multiband character of superconductivity and several 
competing pairing channels.

Experimental data suggest that in the hole-doped 122 compounds 
Ba$_{1-x}$K$_x$Fe$_2$As$_2$, the symmetry of superconducting state can 
change depending on the doping level $x$. A typical band structure of
Ba$_{1-x}$K$_{x}$Fe$_2$As$_2$ consists of two hole pockets at the
$\Gamma$ point and two electron pockets at $(0,\pi)$ and $(\pi,
0)$. At moderate doping level $x\sim 0.4$ various measurements, including 
ARPES \cite{Ding.Richard.ea:08,Khasanov.Evtushinsky.ea:09,Nakayama.Sato.ea:11}, 
thermal conductivity \cite{Luo.Tanatar.ea:09} and neutron scattering 
experiments \cite{Christianson.Goremychkin.ea:08}, are consistent with 
the hypothesis of an $s_\pm$ state where the superconducting gap changes 
sign between electron and hole pockets.
On the other hand, the symmetry of the superconducting state at heavy doping 
$x\rightarrow 1$ is not so clear regarding the question whether the $d$ 
channel dominates or if the gap retains $s_\pm$-symmetry changing sign 
between the inner hole bands at the $\Gamma$ point \cite{KorshunovSwave1,
KorshunovSwave2}. Indeed, there are evidences that $d$-wave pairing
channel dominates \cite{Reid.Juneau-Fecteau.ea:12,ExperimentsDiS1,
ExperimentsDiS2,ExperimentsDiS3} while other ARPES data were interpreted
in favour of an $s$-wave symmetry \cite{SWaveHoleDoped1,SWaveHoleDoped2}.
In both situations this implies the possible existence of an intermediate 
complex state that ``compromises" between the behaviours at moderate and 
high doping. Depending on whether $d$ or $s$ channel dominates at heavy 
doping such a complex state is named $s+is$ or $s+id$.

The $s+is$ state is isotropic and preserves crystal symmetries \cite{Chubukov2}. 
On the other hand, the $s+id$ state breaks the $C_4$ symmetry, while it remains 
invariant under combination of time-reversal symmetry operation and $C_4$ rotations. 
Being anisotropic, it is thus qualitatively different from $s+is$ state. 
Note that the $s+id$ superconducting state is also qualitatively different 
from the (time-reversal preserving) $s+d$ states that attracted interest in 
the context of high-temperature cuprate superconductors (see e.g. \cite{Joynt:90,
Li.Koltenbah.ea:93,Berlinsky.Fetter.ea:95}). 
It also contrasts with $d+id$ state, which can appear in the presence of an 
external magnetic field, and that violates both parity and time-reversal 
symmetries \cite{Balatsky:00, Laughlin:98} . 
While it is an interesting scenario, possibly relevant for pnictides, 
we will not further consider the properties of $s+id$ state and focus on a 
detailed analysis of $s+is$ superconducting state. This state is indeed of 
particular theoretical interest, being the simplest BTRS extension of the 
most abundant $s$-wave state. Also, it is expected to arise from various 
microscopic physics \cite{StanevTesanovic,Thomale,Suzuki,Chubukov1,Chubukov2,
ChubukovLiFeAs}. The $s+is$ state could as well be fabricated on demand 
on the interfaces of superconducting bilayers \cite{Ng}. 

To this day, no experimental proof of $s+is$ nor $s+id$ BTRS states have been 
reported. Indeed this requires probing the relative phases between the components 
of the order parameter in different bands, which is a challenging task.
For example the $s+is$ state does not break the point group symmetries and 
is therefore not associated with an intrinsic angular momentum of Cooper pairs. 
Consequently it cannot produce a local magnetic field and thus is {\it a priori} 
invisible for conventional methods like muon spin relaxation and polar Kerr 
effect measurements that were for example used to probe time-reversal breaking 
$p+ip$ superconducting state in e.g. Sr$_2$RuO$_4$ compound \cite{Mackenzie.Maeno:03}. 
Several proposals have been recently voiced, each with various limitations, 
for indirect observation of BTRS signatures in pnictides. These, for example, 
include the investigation of the spectrum of the collective modes which includes 
massless \cite{Lin} and mixed phase-density \cite{Johan,Stanev,Chubukov2,Benfatto} 
excitations.
Also, it was proposed to consider the properties of exotic topological excitations 
such as skyrmions and domain walls \cite{Garaud.Carlstrom.ea:11,Garaud.Carlstrom.ea:13,
Garaud.Babaev:14}, unconventional mechanism of vortex viscosity \cite{Silaevvisc}, 
formation of vortex clusters \cite{Johan}, exotic reentrant and precursor phases 
induced by fluctuations \cite{bojesen2013time,bojesen2014phase,carlstrom2014spontaneous,
hinojosa2014time}.
Spontaneous currents were predicted to exist near non-magnetic impurities in 
anisotropic superconducting $s+id$ states \cite{Lee.Zhang.Wu:09,ChubukovMaitiSigrist} 
or in samples subjected to strain \cite{ChubukovMaitiSigrist}. The latter proposal 
actually involves symmetry change of $s+is$ states and relies on the presence of 
disorder which can typically have uncontrollable distribution.
It was also recently pointed out that the time-reversal symmetry breaking $s+is$ 
state features an unconventional contribution to the thermoelectric effect 
\cite{Silaev.Garaud.ea:15}. Related to this an experimental set-up, based on a local 
heating was recently proposed \cite{Garaud.Silaev.ea:15}. The key idea being that 
local heating induces local variations of relative phases which further yield 
an electromagnetic response that is directly observable. 

The paper is organized as follows: in section \ref{Sec:Microscopic} we start by 
deriving the GL expansion for a generic $N$-band model. Then we focus on the 
minimal three-band microscopic model suggested to describe hole-doped 122 compounds 
with three superconducting gaps. There we consistently derive the two-component 
Ginzburg-Landau equations that are relevant for interband dominated pairing.
Next, section \ref{Sec:GLgroundstate} is devoted to the analysis of the ground-state 
phases of the Ginzburg-Landau model. Then in section \ref{Sec:GLproperties} we 
introduce a reparametrization of the model that allows further investigation of 
the perturbative spectrum. This allows for example to derive the relevant length 
scales and the second critical field. Finally, section \ref{Ref:Topology} is devoted 
to the analysis of the topological properties of the theory, together with the 
possible topological excitations.

\section{Microscopic model and derivation of the multi-component Ginzburg-Landau expansion} 
\label{Sec:Microscopic}

\begin{figure}[!tb]
\hbox to \linewidth{\hss
\includegraphics[width=.75\linewidth]{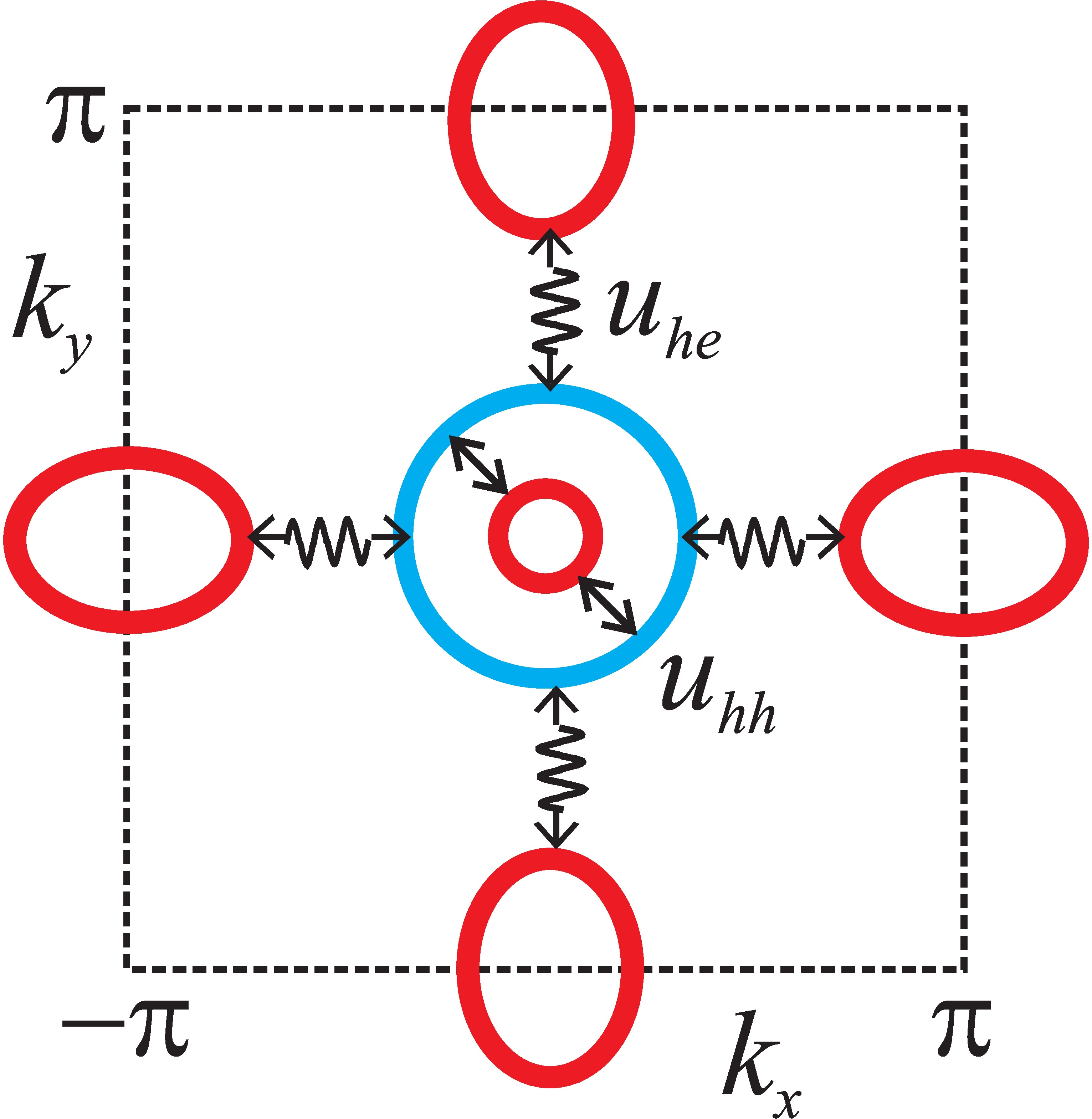}
\hss}
\caption{
(Color online) --
Schematic view of the band structure in hole-doped iron pnictide compound 
Ba$_{1-x}$K$_x$Fe$_2$As$_2$. It consists of two hole pockets at $\Gamma$ 
shown by circles and two electron pockets at $(0; \pi)$ and $(\pi; 0)$ 
shown by ellipses. As discussed in the text, the $s+is$ state is favoured 
by superconducting coupling dominated by the interband repulsion between 
electron and hole Fermi surfaces $u_{he}$, as well as between the two hole 
pockets $u_{hh}$.
}
\label{Fig:BZ}
\end{figure}

We consider superconducting coupling which can result on BTRS state. A typical 
band structure of iron pnictides consists of two hole pockets at the $\Gamma$ 
point and two electron pockets at $(0,\pi)$ and $(\pi,0)$. This structure is 
sketched on \Figref{Fig:BZ}, where the dominating pairing channels are the 
interband repulsion between electron and hole bands, as well as between the two 
hole pockets at $\Gamma$. Note that there, the order parameter is the same in 
both electron pockets so that the crystalline $C_4$ symmetry is not broken and 
thus corresponds to an $s$ state. 
This contrasts with an alternative scenario which was proposed in pnictides, the 
strongest interactions are the repulsions between hole and electron bands and 
between two electron pockets. Such interaction favours order parameter sign change 
between electron pockets resulting in a $C_4$ symmetry breaking $d$-wave state. 
Such a scenario also allows BTRS phase in the form of an $s+id$ superconducting 
state. As symmetrywise it is qualitatively different from the $s+is$ state, 
the case of an $s+id$ superconducting state is not discussed here.

\subsection{Generic \texorpdfstring{$N$}{N}-component expansion}

To derive a Ginzburg-Landau expansion that can be used for example in 
numerical simulations, we consider the microscopic model of clean superconductor 
with, in general, $N$ overlapping bands at the Fermi level. Within the 
quasiclassical approximation, the band parameters that characterize the 
different cylindrical sheets of the Fermi surface are the Fermi velocities 
${\bm v}^{(j)}_{F}$ and the partial densities of states (DOS) $\nu_j$, where 
the label $j=1,...,N$ denotes the band index. The particular example of 
a three-band model believed to be relevant in 122 compounds and which is 
considered in details below, is schematically shown in \Figref{Fig:BZ}. 
The Eilenberger equations for quasiclassical propagators read as
\SubAlign{Eq:EilenbergerF}{
 &\hbar{\bm v}^{(j)}_{F}{\bm \Pi}\phantom{^*} {f^{\phantom{+}}_j} +
 2\omega_n f_j^{\phantom{+}} - 2 \Delta_j g_j=0, \\ 
 &\hbar{\bm v}^{(j)}_{F}{\bm \Pi}^* f^+_j -
 2\omega_n f^+_j + 2\Delta^*_j g_j=0 \,,
}
where ${\bm \Pi}=\Grad - 2\pi i \A/\Phi_0$, $\A$ is the vector potential, and 
$\Phi_0$ the flux quantum. The quasiclassical Green's functions in each band obey 
the normalization condition $g_j^2+f_jf_j^+=1$. Finally, the self-consistency 
equation for the gaps and the electric current are
\Align{}{
 \Delta_i ({\bm p},{\bm r})&= 2\pi T \sum_{n,{\bm p^\prime},j} 
 \lambda_{ij}({\bm p},{\bm p^\prime}) f_j ({\bm p},{\bm r}, \omega_n) 
 \label{Eq:SelfConsistentGap} ,\,	\\ 
 {\bm j} ({\bm r})&=  2\pi e T  \sum_{n, {\bm p},j}
 \nu_j{\bm v}_F^{(j)} \Im\; g_j ({\bm p},{\bm r}, \omega_n)
 \label{Eq:SelfConsistentCurrent}\,.
}
Here $g_j=\mathrm{sign} (\omega_n)\sqrt{1- f_jf^+_j}$ and $\nu_j$ is the density 
of states, the parameters ${\bm p}$ run over the corresponding Fermi surfaces 
and $\lambda_{ij}$ are the components of the coupling potential matrix. For 
simplicity we further consider isotropic pairing states so that  
$\lambda_{ij}({\bm p},{\bm p^\prime} )= const$, on each of the Fermi surfaces. 
However, in electron pockets we keep the anisotropy of Fermi velocities in 
Eqs.\Eqref{Eq:EilenbergerF}. The anisotropy of hole bands on the other hand can 
be neglected, which is a well-justified assumption \cite{VorontsovVekhter}.

The derivation of the Ginzburg-Landau functional from the microscopic equations 
formally follows the standard scheme. That is by expressing the solutions of the 
Eilenberger equations in the form of the expansion by powers of the gap functions 
amplitudes $\Delta_{j}$ and their gradients. We stress that this multiband expansion 
relies on the existence of above-mentioned multiple small parameters. This is in 
contrast to the simplest single-band Ginzburg-Landau expansion that is justified 
by a single small parameter $(1- T/T_c)$. For a multi-band GL expansion, using a 
single small parameter is in general incorrect or irrelevant due to the existence of 
multiple broken symmetries. Also, even in the case of a multiband system breaking a 
single symmetry, an expansion that is obtained relying on a single small parameter in 
certain cases has vanishingly small applicability range \cite{Silaev.Babaev:12}. 
For earlier works on microscopic two-band GL expansions see \cite{Tilley:64,Gurevich:03,
Zhitomirsky.Dao:04,Gurevich:07} and the analysis is straightforwardly generalized 
to similar $s$-wave $N$-component cases.
First, the solutions of Eqs.\Eqref{Eq:EilenbergerF} are found in the form of the 
expansion by powers of the gap functions amplitudes $\Delta_{j}$ and their gradients: 
\Align{Eq:GLexpansion}{
 & f_j({\bm p},{\bm r}, \omega_n) = \\ \nonumber
 \frac{\Delta_j}{\omega_n}-\frac{|\Delta_j|^2\Delta_j}{2\omega_n^3}
 &-\frac{ \hbar( {\bm v}^{(j)}_{F} {\bm \Pi}) \Delta_j }{{2\omega_n^2}}
 +\frac{\hbar^2( {\bm v}^{(j)}_{F} {\bm \Pi}) 
 ( {\bm v}^{(j)}_{F} {\bm \Pi}) \Delta_j}{4\omega_n^3}
}
and $ f^+_j ({\bm p},{\bm r}, \omega_n)  = f^*_j(-{\bm p},{\bm r}, \omega_n)$.
Then, the summation over Matsubara frequencies gives
\Equation{}{
 2\pi T \sum_{n=0}^{N_d} \omega_n^{-1} = G_0 + \tau \,,
 ~~\text{with}~~\tau = (1- T/T_c)\,.
}
Substituting \Eqref{Eq:GLexpansion} into the self-consistency 
Eqs.\Eqref{Eq:SelfConsistentGap} and normalizing the gaps by $T_c/\sqrt{\rho}$ 
(where $\rho =\sum_n \pi T_c^3\omega_n^{-3} \approx 0.1$), determines the system of 
GL equations
\Equation{Eq:GL3Band}{
  \big[(G_0+\tau-\hat\Lambda^{-1}){\bm\Delta} \big]_j
  = -K^{(j)}_{ab}\Pi_a\Pi_b \Delta_j + |\Delta_j|^2\Delta_j  \,,
}
where
${\bm \Delta} = (\Delta_1,\cdots,\Delta_N)^T$. The anisotropy tensor is 
$K^{(j)}_{ab} =\hbar^2\rho \langle v^{(j)}_{Fa}v^{(j)}_{Fb} \rangle /2T_c^2$, 
where the average is taken over the $j$-th Fermi surface and $a,b$ stand for 
the $x,y$ coordinates.
The current reads as
 \Equation{Eq:CurrentGL}{
 {\bm J} ({\bm r})= \frac{4e}{\hbar}\frac{T_c^2}{\rho}
 \sum_{i=1}^N \nu_i \Im\; \Delta^*_i \hat K_i {\bm \Pi} \Delta_i \,.
}

The critical temperature is determined by the equation 
$G_0 = \min_n (\lambda^{-1}_n)$, where $\lambda_{n}^{-1}$ are the positive 
eigenvalues of the inverse coupling matrix $\hat\Lambda^{-1} $. 
Provided that all the eigenvalues are positive, the number of components of 
the order parameter coincide with the number of bands $N$. In this case, the 
system of Ginzburg-Landau equations for the general $N$-component system 
can be written as follows 
\Equation{Eq:GLNcomponent}{
   -K^{(i)}_{ab}\Pi_a\Pi_b \Delta_i + \alpha_i\Delta_i 
   + \eta_{ij}\Delta_j + \beta_i|\Delta_i|^2\Delta_i =0,
}
where 
\SubAlign{Eq:GLNcomponentCoef}{ 
\alpha_i = (\hat\Lambda^{-1}_{ij} - G_0-\tau)\delta_{ij}, \\
 \eta_{ij} = (1-\delta_{ij}) \hat\Lambda^{-1}_{ij} 
 \;\;\; {\rm and} \;\;\; \beta_i = 1 \,.
} 

Various choices of microscopic coupling parameters can result in qualitatively 
different structures of the Ginzburg-Landau field theory. 

\subsection{Ginzburg-Landau models for the \texorpdfstring{$s+is$}{s+is} state}

Our main interest here is the time-reversal symmetry breaking $s+is$ states in 
three-band systems. Let us consider first the simplest case of an intraband dominated 
pairing which can be described by a three-component GL theory in the vicinity of $T_c$. 
This regime is described by the following coupling matrix
\begin{equation}\label{Eq:Model3Band1}
\hat\Lambda =  \left(%
\begin{array}{ccc}
      \lambda & -\eta_h & -\eta_e \\
      -\eta_h & \lambda & -\eta_e \\
      -\eta_e & -\eta_e & \lambda
\end{array}%
\right)	\,,
\end{equation}
where $\eta_e, \eta_h \ll \lambda$. The critical temperature is
determined by the equation $G_0 = \min (\lambda^{-1}_1, \lambda^{-1}_2,
\lambda^{-1}_3)$, where $\lambda_{1,2}^{-1} = (2\lambda-\eta_h \pm
\sqrt{8\eta_e^2+\eta_h^2})/[2(\lambda^2 - \lambda\eta_h-2\eta_e^2)] $
and  $\lambda_{3}^{-1} = 1/(\lambda+\eta_h) $ are the positive eigenvalues
of the inverse coupling matrix
\begin{equation}\label{Eq:Model3Band2}
\hat\Lambda^{-1} =  X\left(%
\begin{array}{ccc}
    \lambda^2- \eta_e^2    & \eta_e^2+\lambda\eta_h  & \eta_e(\lambda+\eta_h) \\
    \eta_e^2+\lambda\eta_h & \lambda^2- \eta_e^2    & \eta_e(\lambda+\eta_h) \\
    \eta_e(\lambda+\eta_h) & \eta_e(\lambda+\eta_h)     & \lambda^2- \eta_h^2
\end{array}%
\right)		\,,
\end{equation}
where $X=1/[(\lambda^2- \lambda\eta_h- 2\eta_e^2)(\lambda+\eta_h)]$.
Since we assume that $\eta_{e,h} >0 $ and $\eta_{e,h} \ll \lambda $, 
the critical temperature is given by the smallest eigenvalue
$G_0 = 1/(\lambda+\eta_h) $ so that
\begin{equation}
G_0 - \hat\Lambda^{-1} =  - \left(%
\begin{array}{ccc}
  a_1 & a_1 & a_2 \\
  a_1 &  a_1 & a_2 \\
  a_2 & a_2 &  a_3 \\
\end{array}%
\right) \,,
\end{equation}
where
\begin{subequations} \label{Eq:GLcoeff}
\begin{eqnarray}
   a_1 &=& (\eta_e^2+\lambda\eta_h)/X \\
   a_2 &=& \eta_e(\lambda+\eta_h)/X\\
   a_3 &=& (2\eta_e^2-\eta_h^2+\lambda\eta_h)/X\,.
\end{eqnarray}
\end{subequations}
The free energy functional whose variations give the three-component 
Ginzburg-Landau equations reads as 
\Align{Eq:EnergyExample}{
\F=\frac{\B^2}{2} &+\sum_{j=1}^3\Big\{
    \frac{k_j}{2}\left|\bm{\Pi}\Delta_j\right|^2
    +\alpha_j|\Delta_j|^2+ \frac{\beta_j}{2}|\Delta_j|^4  \Big\}
    \nonumber \\
    &+\sum_{j=1}^3\sum_{k=j+1}^3\eta_{jk}\Big\{
    \Delta_j^* \Delta_k+\Delta_k^* \Delta_j   \Big\}  \,,
 }
where $\beta_k=1$, $\eta_{12}=a_1$, $\eta_{13}=\eta_{23}=a_2$, 
$\alpha_k = \alpha_k^0 (T/T_k -1)$, $\alpha_k^0= 1-a_k$, $T_k= T_c (1-a_k)$. 
Here we assumed that bands are isotropic and put 
$K^{(j)}_{xx}= K^{(j)}_{yy}= k_j\xi_0^{2}/2$ and introduce the dimensional 
units normalizing lengths by $ \xi_0 = \hbar \bar{v}_F/T_c$ (where $\bar{v}_F$ 
is the average value of Fermi velocity), magnetic field by 
$B_0 = T_c\sqrt{4\pi\nu/\rho}$, current density $j_0=c B_0/(4\pi\xi_0)$ and 
free energy density $F_0= B_{0}^2/4\pi$. Here $\nu$ is the DOS which is assumed 
to be the same in all superconducting bands.  In such units the electric charge 
is replaced by an effective coupling constant $\tilde{e} = 2\pi B_0 \xi_0^2/\Phi_0$ 
so that $\bm{\Pi}= \nabla+ i\tilde{e}{\bm  A}$. Below we omit the tilde for brevity.


In the following, we consider another choice of the coupling matrix 
$\hat{\Lambda}$, suggested to be relevant for iron-based superconductors 
\cite{Chubukov1,Chubukov2,Benfatto}. It corresponds to the case of an 
interband dominated pairing with repulsion, parametrized as
\begin{equation}\label{Eq:Model3BandB1}
\hat\Lambda = - \left(%
\begin{array}{ccc}
  0        & \eta    & \lambda   \\
  \eta     & 0       & \lambda   \\
  \lambda  & \lambda & 0         \\
\end{array}  \right) \,.
\end{equation}
Here $\Delta_{1,2}$ correspond to the gaps at hole Fermi surfaces and $\Delta_3$ 
is the gap at electron pockets so that the coefficients $\eta= u_{hh}$ and
$\lambda=u_{eh}$ are respectively the hole-hole and electron-hole interactions. 
Neglecting the r.h.s. in \Eqref{Eq:GL3Band} we get the linear equation which 
determines the critical temperature $G_0 =\min\left(G_1, G_2\right)$, where 
$G_1 = 1/\eta$ and $G_2 =\left(\eta+\sqrt{\eta^2+8\lambda^2} \right) /4\lambda^2$
are the positive eigenvalues of the matrix
\Equation{Eq:Model3BandB2}{
 \hat\Lambda^{-1} =  \frac{1}{2\lambda^2\eta}\left(
 \begin{array}{ccc}
  \lambda^2     & -\lambda^2    & -\lambda\eta  \\
  -\lambda^2    & \lambda^2     & -\lambda\eta  \\
  -\lambda\eta  & -\lambda\eta  & \eta^2        \\
\end{array}  \right) \,.
}
The coupling matrix $\hat\Lambda^{-1}$ has only two positive eigenvalues $G_{1,2}$ 
whose eigenvectors are ${\bm\Delta}_1=(-1,1,0)^T$ and ${\bm\Delta}_2=(x,x,1)^T$ 
with $x=(\eta-\sqrt{\eta^2+8\lambda^2})/4\lambda$. Since only the fields 
corresponding to the positive eigenvalues can nucleate, the GL theory 
\Eqref{Eq:GL3Band} has to be reduced to a two-component one. 
This reduction is 
obtained by expressing the general order parameter in terms of the superposition
\Align{Eq:OPreotation}{
 {\bm \Delta}&=  \psi_1 {\bm \Delta}_1 + \psi_2 {\bm \Delta}_2\,, \nonumber \\
 \text{and}~~~(\Delta_1,\Delta_2,\Delta_3)&=(x\psi_2-\psi_1,x\psi_2+\psi_1,\psi_2)\,.
}
There, $\psi_1$ and $\psi_2$ are the order parameter of $s_\pm$ pairing channels 
between two concentric hole surfaces and between hole and electron surfaces 
correspondingly.

Now, substituting the ansatz \Eqref{Eq:OPreotation} into the system of 
Ginzburg-Landau equations \Eqref{Eq:GL3Band} we obtain, after projection 
onto the vectors ${\bm \Delta}_{1,2}$, the system of two GL equations
\SubAlign{Eq:GL3BandReduced}{ 
 a_1\psi_1 + b_{1j}|\psi_j|^2\psi_1+b_{J}\psi_1^*\psi_2^2 &= \\  
 ( K_{aa}^{(1)} + K_{aa}^{(2)} ) \Pi_a^2 \psi_1 
 &+ x ( K_{aa}^{(2)} - K_{aa}^{(1)} ) \Pi_a^2  \psi_2  \nonumber \\ 
   a_2\psi_2 + b_{2j}|\psi_j|^2\psi_2+b_{J}\psi_2^*\psi_1^2 &= \\
 \left[ x^2( K_{aa}^{(1)} + K_{aa}^{(2)} ) + K_{aa}^{(3)} \right] \Pi_a^2 \psi_2 
 &+ x ( K_{aa}^{(2)} - K_{aa}^{(1)} ) \Pi_a^2 \psi_1.  \nonumber
}
The parameters of the left hand side of the Ginzburg-Landau equations 
\Eqref{Eq:GL3BandReduced} are expressed, in terms of the coefficients 
of the coupling matrix \Eqref{Eq:Model3BandB1} as 
\SubAlign{Eq:Parameters}{
a_j &= -|\bm \Delta_j|^2(G_0-G_{j}+\tau )\,,~~ \\
&~~\text{with}~~ |\bm \Delta_1|^2  =2~~\text{and}~~ |\bm \Delta_2|^2=2x^2+1  \\
b_{11} &= 2    \,,~~~b_{22}=(2x^4+1) ~~\text{and}~~b_k:=b_{kk}\\
b_{12} &= 4x^2 \,,~b_J=2x^2 \,.
}
As previously stated, the $s+is$ state is symmetric under $C_4$ transformations, 
which implies that $K^{(j)}_{xx} = K^{(j)}_{yy} = K^{(j)}$. 
The Ginzburg-Landau equations can thus be further simplified as follows
\SubAlign{Eq:GL3BandReducedSiS}{
 a_1\psi_1 + b_{1j}|\psi_j|^2\psi_1 + b_{J}\psi_1^*\psi_2^2 & = 
\frac{k_{1j}}{2} {\bm \Pi}^2  \psi_j  \,,
\\
 a_2\psi_2 + b_{2j}|\psi_j|^2\psi_2 + b_{J}\psi_2^*\psi_1^2 &= 
\frac{k_{2j}}{2} {\bm \Pi}^2  \psi_j  \,,
}
where the coefficients of the gradient terms read as 
\SubAlign{Eq:GradientCoeffitients}{
   k_{11}:=k_1 & = 2\xi_0^{-2}\big[ K^{(1)} +  K^{(2)} \big]				\\
   k_{22}:=k_2 & = 2\xi_0^{-2}\big[(K^{(1)} + K^{(2)})x^2 + K^{(3)} \big]	\\
   k_{12}      & = 2\xi_0^{-2} x\big[K^{(2)} - K^{(1)} \big] \,.
   }
The total current (\ref{Eq:CurrentGL}), which can be expressed in terms 
of the partial currents $\J^{(i)}$ of the different components of order 
parameters, reads as $\J=\sum_i\J^{(i)}$; and the partial currents read as 
\Equation{Eq:Currents}{
\J^{(i)} = e\Im\big(\psi_i^*\left[k_i\D\psi_i+k_{12}\D\psi_j\right]\big)
\,,
}
where $j\neq i$. 

The two-component free energy functional that corresponds to the 
Ginzburg-Landau equations \Eqref{Eq:GL3BandReducedSiS}, and whose variations 
with respect to $\A$ give the supercurrent \Eqref{Eq:Currents}, reads as 
(in dimensionless units): 
\SubAlign{Eq:FreeEnergy}{
\F &=\frac{\B^2}{2}+\sum_{j=1}^2\Big\{
 	\frac{k_{j}}{2}\left|\D\psi_j \right|^2
 	+a_j|\psi_j|^2+\frac{b_{j}}{2}|\psi_j|^4\Big\}  
 	\label{Eq:FreeEnergy:Self}	\\
   &+\frac{k_{12}}{2}
	\Big((\D\psi_1)^*\D\psi_2+(\D\psi_2)^*\D\psi_1 \Big)
	\label{Eq:FreeEnergy:Mixed}	\\
   &+b_{12}|\psi_1|^2|\psi_2|^2
	+\frac{b_J}{2}\big(\psi_1^{*2}\psi_2^2+c.c.\big)
	\label{Eq:FreeEnergy:Interaction}	\,.
}
Here, the complex fields $\psi_j=|\psi_j| e^{i\theta_j}$ represent the 
components of the order parameter. These are electromagnetically coupled 
by the vector potential $\A$ of the magnetic field $\B=\bs\nabla\times\A$, 
through the gauge derivative $\D\equiv\Grad+ie\A$ where the coupling 
constant $e$ is used to parametrize the London penetration length.
Note that for the energy to be positively defined, the coefficients of 
the kinetic terms should satisfy the relation $k_1k_2-k_{12}^2>0$. Also, 
for the free energy functional to be bounded from below, the coefficients 
of the terms that are fourth order in the condensates should satisfy the 
constraint $b_1b_2-(b_{12}+b_J)^2>0$. These conditions are of course 
satisfied by the microscopically calculated value \Eqref{Eq:Parameters} 
and \Eqref{Eq:GradientCoeffitients}.

Note that the three-band model considered considered here can describe the 
formation of $s+id$ state due to the competition between electron-electron 
and electron-hole repulsion. In this case one can derive a two-component 
Ginzburg-Landau model in the same line as described above \cite{Garaud.Silaev.ea:15}. 
The only qualitative difference between $s+is$ and $s+id$ states is contained 
in the structure of mixed gradient term, which is isotropic in the former case 
while in the latter changes sign due to the C$_4$ rotation.  


\section{Ground-state properties of the two-component Ginzburg-Landau model} 
\label{Sec:GLgroundstate}

Depending on the relation between the parameters of the potential, qualitatively 
different superconducting phases can be identified. These are determined by the 
ground-state properties of the theory. Since the coefficients of the kinetic terms 
satisfy the relation $k_1k_2-k_{12}^2>0$, the ground state is homogeneous ($\D\psi_k=0$) 
and thus it is determined only from the potential terms of \Eqref{Eq:FreeEnergy} 
that reads as: 
\Align{Eq:Potential}{
V=\sum_{j=1}^2 &a_j|\psi_j|^2+\frac{b_j}{2}|\psi_j|^4 \nonumber \\
&+ |\psi_1|^2|\psi_1|^2\big(b_{12}+b_J\cos\theta_{12}\big)\,,
}
where $\theta_{12}=\theta_2-\theta_1$ is the relative phase between both 
condensates. The ground state is the state denoted by $\psi_k=u_k\Exp{i\theta_{k}}$, 
and where the vector potential is a pure gauge ($\A=\Grad\chi$ for arbitrary $\chi$) 
that can consistently chosen to be zero. 
The ground state $[u_1,u_2,\theta_{12}]$ is an extremum of that potential 
($\partial V/\partial u_j=0$ and $\partial V/\partial\theta_{12}=0$) 
and thus satisfies: 
\begin{subequations}\label{Eq:Extrema}
\begin{empheq}[left=\empheqlbrace]{alignat=1}
2\left(a_1+b_1u_1^2+(b_{12}+b_J\cos2\theta_{12}) u_2^2\right)u_1 &=0 
 \label{Eq:Extrema1}\\
2\left(a_2+b_2u_2^2+(b_{12}+b_J\cos2\theta_{12}) u_1^2\right)u_2 &=0 
 \label{Eq:Extrema2} \\
-2b_Ju_1^2u_2^2\sin2\theta_{12} &=0
\,. \label{Eq:Extrema3} 
\end{empheq} 
\end{subequations}
Besides being an extremum according to the condition \Eqref{Eq:Extrema}, 
the ground state should also be a minimum. That is, all the eigenvalues of 
the Hessian matrix must be positive. The Hessian matrix reads as 
\begin{widetext}
\Equation{Eq:Hessian}{
{\cal H}=2\left(\begin{array}{ccc}
a_1+3b_1u_1^2 +(b_{12}+b_J\cos2\theta_{12}) u_2^2	& 
2(b_{12}+b_J\cos2\theta_{12}) u_1u_2					& 
-2b_Ju_1u_2^2\sin2\theta_{12}						\\ 
2(b_{12}+b_J\cos2\theta_{12}) u_1u_2					& 
a_2+3b_2u_2^2 +(b_{12}+b_J\cos2\theta_{12}) u_1^2 	& 
-2b_Ju_1^2u_2\sin2\theta_{12}\\
-2b_Ju_1u_2^2\sin2\theta_{12}						& 
-2b_Ju_1^2u_2\sin2\theta_{12}						&
-2b_Ju_1^2u_2^2\cos2\theta_{12}
\end{array}\right)
}
\end{widetext}
Besides the normal state $[u_1,u_2,\theta_{12}]=[0,0,\theta_{12}]$, 
the model \Eqref{Eq:FreeEnergy} features various ground state phases, 
depending on the parameters of the theory. 

\subsection{Normal state}

By definition, the normal state is the state with no superconducting 
condensate: $[u_1,u_2,\theta_{12}]=[0,0,\theta_{12}]$. The Hessian matrix
\Eqref{Eq:Hessian} for that state reads as 
\Equation{Eq:HessianNormal}{
{\cal H}=2\left(\begin{array}{ccc}
a_1 & 0	& 0	\\ 
0	& a_2	& 0\\
0   & 0		& 0
\end{array}\right)\,.
}
It gives the minimal stability condition $a_1,a_2>0$ of the normal state.

\subsection{Phase-separated phase}

\begin{figure*}[!htb]
\hbox to \linewidth{\hss
\includegraphics[width=0.7\linewidth]{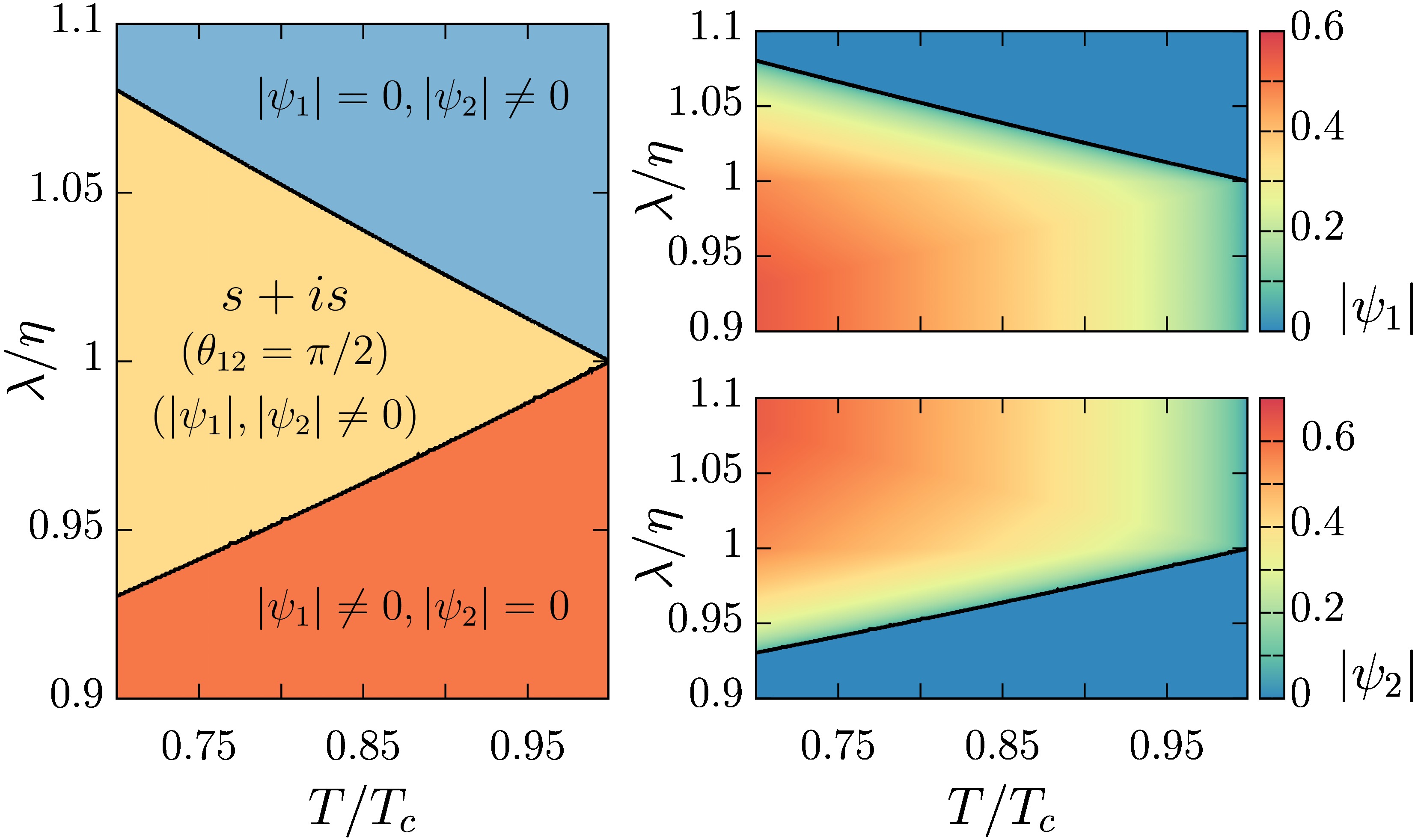}
\hss}
\caption{ (Color online) --
Left panel displays the different phases. In the $s+is$ state, the relative 
phase is $\pi/2$. The panels on the right show the ground state densities 
$|\psi_1|$ and $|\psi_2|$. 
All parameters of the GL functional were calculated using the relations 
\Eqref{Eq:Parameters} where we used $\eta=0.5$.
} \label{Fig:GroundStatePsi}
\end{figure*}

Here, the term \emph{phase-separated phase} relates to the case where only \emph{one}
of Ginzburg-Landau component assumes a non-zero ground state density, while the 
second is completely suppressed. That is, either 
$[u_1,u_2,\theta_{12}]=[\sqrt{-a_1/b_1},0,\theta_{12}]$ 
or $[0,\sqrt{-a_2/b_2},\theta_{12}]$. For example, if only the first component 
has a non-zero ground-state density, the Hessian reads as 
\Equation{Eq:HessianSeparated}{
{\cal H}=2\left(\begin{array}{ccc}
-2a_1 & 0	& 0	\\ 
0	& a_2 +(b_{12}+b_J\cos2\theta_{12}) u_1^2 	& 0\\
0   & 0		& 0
\end{array}\right)\,.
}
Note that the case where only the second component is nonzero can easily be 
obtained by permuting ``1,2" indices. These ground states correspond, in the 
microscopic 3-band model, to two physically different states. The case $u_1\neq0$ 
and $u_2=0$ gives the sign changing gap ($s_\pm$) with $\Delta_3=0$. On the 
other hand, the case $u_2\neq0$ and $u_1=0$ gives the $s_{++}$ state.

\subsection{Coexisting phase}

This is the phase were both $u_1,u_2\neq0$, that we are principally 
interested in, in this paper. Observe that Eq.~\Eqref{Eq:Extrema3} 
specifies the ground state relative phase between both condensates 
to be an integer multiple of $\pi/2$, when both condensates have 
nonzero density $u_1,u_2\neq0$.  Introducing for convenience 
$d=b_{12}+b_J\cos n\pi$, and since the $b_1b_2-d^2>0$, 
for the free energy functional to be bounded from below, the 
ground state is 
\Equation{Eq:GSCoexist}{
[u_1,u_2,\theta_{12}]=
\left[
\sqrt{\frac{a_2d-a_1b_2}{b_1b_2-d^2}}
\,,\sqrt{\frac{a_1d-a_2b_1}{b_1b_2-d^2}	}
\,,\frac{n\pi}{2}\right]
}
and the Hessian matrix becomes
\Equation{Eq:HessianCoexist}{
{\cal H}=4\left(\begin{array}{ccc}
b_1u_1^2 	& 
d u_1u_2					& 
0						\\ 
d u_1u_2					& 
b_2u_2^2 	& 
0\\
0						& 
0						&
-2b_Ju_1^2u_2^2\cos n\pi
\end{array}\right)\,.
}
Obviously, if $b_J<0$ (resp. $b_J>0$) then the stable state has $n$ 
which an even (resp. odd) integer and thus the ground state relative 
phase is $\pm\pi/2$ (resp $0,\pi$) and thus $d=b_{12}+|b_J|$. The 
condition for the stability of the coexisting phase thus boils down 
to having positive eigenvalues for the reduced Hessian
\Equation{Eq:HessianCoexistReduced}{
{\cal H}=4\left(\begin{array}{ccc}
b_1u_1^2 	& 
(b_{12}+|b_J|)u_1u_2					\\ 
(b_{12}+|b_J|)u_1u_2					& 
b_2u_2^2 	
\end{array}\right)\,.
}
 
\subsection{Ginzburg-Landau phase diagram of the \texorpdfstring{$s+is$}{sis} state}

Here we are principally interested in the $s+is$ state. That is in the phase 
where both condensates coexist and the time-reversal symmetry is spontaneously 
broken (i.e. when $\theta=\pm\pi/2$). The ground state thus reads as 
\Equation{Eq:GSsis}{
[u_1,u_2,\theta_{12}]=
\left[
\sqrt{\frac{a_2d-a_1b_2}{b_1b_2-d^2}}
\,,\sqrt{\frac{a_1d-a_2b_1}{b_1b_2-d^2}	}
\,,\pm\frac{\pi}{2}\right]\,,
}
in the region of the parameter space defined by 
\Align{Eq:GSsisStability}{
b_1b_2-(b_{12}+b_J)^2>&0 \\
a_2d-a_1b_2>&0 \\
a_1d-a_2b_1>&0 \,.
}
  
The physical system we are interested in here, is the microscopic three-band 
model \Eqref{Eq:Model3BandB1} with interband dominated pairing with repulsion, 
that is believed to be relevant for some pnictides. There, the parameters of 
the Ginzburg-Landau model \Eqref{Eq:FreeEnergy} actually depend on few microscopic 
parameters of the coupling matrix and the reduced temperature, as shown on 
\Eqref{Eq:Parameters}. \Figref{Fig:GroundStatePsi} shows the ground state phase 
diagram of the Ginzburg-Landau model as a function of the reduced temperature 
and the ratio of electron-hole and hole-hole interactions.

\section{Elimination of mixed gradients and Separation of charged and neutral modes} 
\label{Sec:GLproperties}

\begin{figure*}[!htb]
\hbox to \linewidth{\hss
\includegraphics[width=0.7\linewidth]{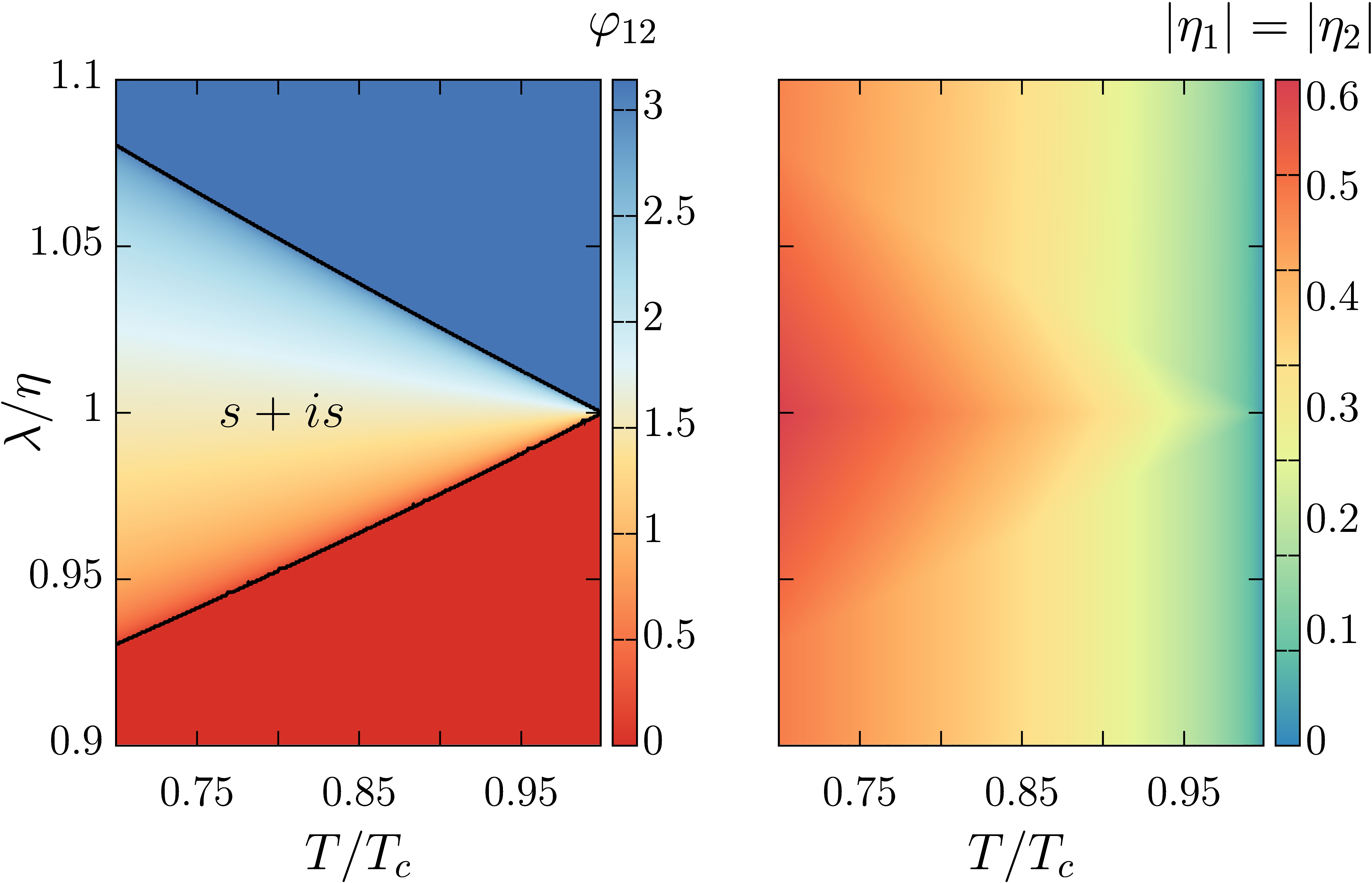}
\hss}
\caption{ (Color online) --
Ground state phase diagram in the $\eta$ parametrization where there are no 
mixed gradients. Left panel displays the different phases. In the $s+is$ state, 
the relative phase is different from 0, $\pi$. The right panel show the ground 
state densities $|\eta_1|=|\eta_2|$. This is for the same values of the microscopic 
parameters as in \Figref{Fig:GroundStatePsi} and for the value of the coefficients 
of the kinetic terms $K^{(1)}=0.5$, $K^{(2)}=0.05$ and $K^{(3)}=0.25$.
} \label{Fig:GroundStateEta}
\end{figure*}

The current basis for the superconducting degrees of freedom is quite convenient, 
as it allows to easily characterize the ground state and its stability properties. 
On the other hand, it is quite complicated to deal with the kinetic terms. 
This is why it is worth rewriting the model using a linear combination of the 
components of the order parameter that diagonalize the the kinetic terms:
\Equation{Eq:NewFields}{
\eta_1=\sqrt{k_1}\psi_1+\sqrt{k_2}\psi_2\,,~~
\eta_2=\sqrt{k_1}\psi_1-\sqrt{k_2}\psi_2\,.
}
Within this new basis, the kinetic term has a much simpler form. The potential, 
on the other hand becomes much more complicated and at first glance it is not 
possible to find the ground state analytically. However, since it is known from 
the old field basis, it is actually quite simple to derive the analytic solution.
It is also a convenient basis to investigate the physical 
length scales, critical fields, as well as describing various unusual properties. 
In the new field basis, the free energy reads as 
\SubAlign{Eq:FreeEnergy:transf}{
&\F=\frac{\B^2}{2}+\sum_{a=1}^2
 	\frac{\kappa_a}{2}\left|\D_j\eta_a \right|^2
 	+\alpha|\eta_a|^2+\frac{\beta}{2}|\eta_a|^4
 	\label{Eq:FreeEnergy:transf:Self}	\\
   	&+\left(\nu+\omega\left(|\eta_1|^2|+\eta_2|^2\right)\right)
   	|\eta_1||\eta_2|\cos\varphi_{12}
   	\label{Eq:FreeEnergy:transf:Interaction1}	\\
	&+\left(\gamma+\delta\cos2\varphi_{12}\right)|\eta_1|^2|\eta_2|^2
	\label{Eq:FreeEnergy:transf:Interaction2}	\,,
}
with $\eta_a=|\eta_a|\Exp{i\varphi_a}$, $\varphi_{12}=\varphi_2-\varphi_1$ and 
the coefficients for the kinetic term are now
\Equation{Eq:CoeffKinetic}{
\kappa_1=\frac{\sqrt{k_1k_2}+ k_{12}}{2\sqrt{k_1k_2}}~~~~\text{and}~~~~
\kappa_2=\frac{\sqrt{k_1k_2}- k_{12}}{2\sqrt{k_1k_2}}
\,.
}
The coefficients of the potential read as
\Align{Eq:Coeffients2}{
\alpha&=	\frac{a_1k_2+a_2k_1}{4k_1k_2}	\nonumber\\
\nu&=\frac{a_1k_2-a_2k_1}{2k_1k_2}		\nonumber\\
\beta&=\frac{b_1k_2^2+b_2k_1^2+2k_1k_2(b_{12}+b_J)}{16k_1^2k_2^2}		\nonumber\\
\omega&=\frac{b_1k_2^2-b_2k_1^2}{8k_1^2k_2^2}		\nonumber\\
\gamma&=\frac{b_1k_2^2+b_2k_1^2-2k_1k_2b_J}{8k_1^2k_2^2}		\nonumber\\
\delta&=\frac{b_1k_2^2+b_2k_1^2+2k_1k_2(b_J-b_{12})}{16k_1^2k_2^2} \,.
}

In the new field basis, the Ginzburg-Landau equations have no mixed gradients and 
read as 
\Equation{Eq:EOM:transf}{
\D^2\eta_i
=2\frac{\partial V}{\partial\eta_i^*}\,,
} 
while variation of the free energy \Eqref{Eq:FreeEnergy:transf} with respect to 
the vector potential $\A$, determines Amp\`ere's equation $\Curl\B+\J=0$. The total 
current is the superposition of the partial currents ($\J=\sum_k\J^{(k)}$) 
that reads as 
\Equation{Eq:Currents:transf}{
\J^{(i)} = e\kappa_i\Im\big(\eta_i^*\D\eta_i\big)\,.
}
This reparametrization simplifies drastically the Ginzburg-Landau equations 
as there is no more  coupling of the components through mixed gradients. 
However, this comes with the price of more complicated potential terms. 
This is actually a minor problem, since the ground state within the new basis, 
can easily be determined from the one in the old basis according to the formulas
\Align{Eq:Transform}{
|\eta_1|^2&=k_1|\psi_1|^2+k_2|\psi_2|^2
	+2\sqrt{k_1k_2}|\psi_1||\psi_2|\cos\theta_{12}	
	\,, \nonumber\\
|\eta_2|^2&=k_1|\psi_1|^2+k_2|\psi_2|^2
	-2\sqrt{k_1k_2}|\psi_1||\psi_2|\cos\theta_{12}	
	\,, \nonumber\\
\varphi_{12}&=\tan^{-1}\left(
\frac{-2\sqrt{k_1k_2}|\psi_1||\psi_2|\sin\theta_{12}}
	 {k_1|\psi_1|^2-k_2|\psi_2|^2}	\right)\,.
}
\Figref{Fig:GroundStateEta} shows the ground state phase diagram expressed in the 
new parametrization \Eqref{Eq:NewFields}. Note that the ground state diagram 
\Figref{Fig:GroundStatePsi} do not depend on the values the components of the 
(microscopic) anisotropy tensor. Since the reparametrization \Eqref{Eq:NewFields} 
explicitly depends on the parameters of the kinetic terms, the diagram 
\Figref{Fig:GroundStateEta} also depends on them. However this dependence can be 
only quantitative, since obviously the phase diagram cannot depend on any particular 
choice of parametrization.

\subsection{Separation of charged and neutral modes}

To understand the role of the fundamental excitations, as well as the fundamental 
length scales of the Ginzburg-Landau free energy \Eqref{Eq:FreeEnergy:transf},
it can be rewritten in terms of \emph{charged} and \emph{neutral} modes 
by expanding the kinetic term in \Eqref{Eq:FreeEnergy:transf:Self} and 
using \Eqref{Eq:Currents:transf}:
\Align{Eq:GLRewritten}{
   &\F= \frac{1}{2}(\Curl \A)^2 + \frac{\J^2}{2e^2\varrho^2} 
   +\sum_{a}\frac{\kappa_a}{2}(\Grad|\eta_a|)^2
    \nonumber\\
	&+\frac{\kappa_1\kappa_2|\eta_1|^2|\eta_2|^2}{2\varrho^2}
	(\Grad\varphi_{12})^2 
	+V(|\eta_1|,|\eta_2|,\varphi_{12})
	\,.
}
Here $\varphi_{12}\equiv\varphi_2-\varphi_1$ stands for the relative phase 
between the condensates. For this rewriting, we used the supercurrent defined 
from the Amp\`ere's equation $\Curl\B+\J=0$, that now reads as
\Align{Eq:Currents2}{
  \J/e&= e\varrho^2 \A+\sum_{a}\kappa_a|\eta_a|^2
  \Grad\varphi_a  \,,\nonumber \\
      &\text{with}~~\varrho^2=\sum_{a}\kappa_a|\eta_a|^2
 \,.
}
As discussed below, this formulation allows better interpretation of the small 
perturbations and their length scales, together with a better understanding of 
the elementary topological excitations of the theory.

\subsection{Coherence lengths and perturbation operator}

\begin{figure*}[!htb]
\hbox to \linewidth{\hss
\includegraphics[width=0.7\linewidth]{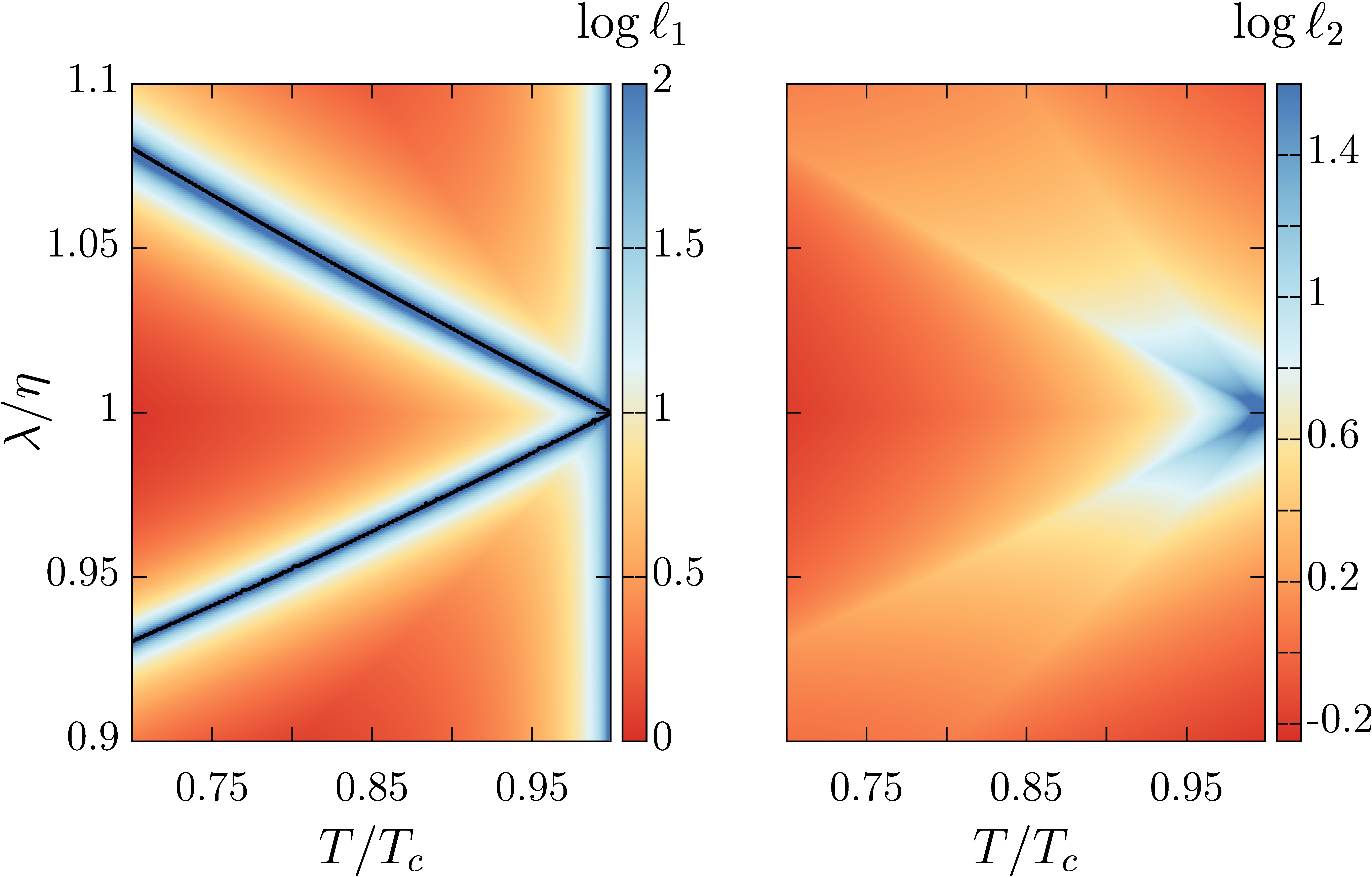}
\hss}
\caption{ (Color online) --
The two largest coherence lengths  (smallest eigenvalues of ${\cal M}^2$). 
The left panel shows the largest coherence length $\ell_1$ that diverges 
at $T_c$, and also when entering the time-reversal symmetry broken 
($s+is$) phase. This indicates a second order phase transition denoted 
by the solid black line.
The right panel displays the second largest length scale $\ell_2$. 
Approaching the point where the system breaks $U(1)\times \groupZ{2}$
symmetry, at mean field level, there should be two divergent coherence 
lengths. Indeed we observe that the second-largest length-scale
$\ell_2$ diverges at a single point at $T_c$ and where $\lambda=\eta$. 
} \label{Fig:GroundLength}
\end{figure*}

The length scales characterizing matter field are the coherence lengths.
These are, by definition, inverse masses of the infinitesimal 
perturbations around the ground state. More precisely, we consider small 
perturbations by linearizing the theory around the ground-state. The eigenspectrum 
of the obtained (linear) differential operator determines the masses of 
the elementary excitations and thus their corresponding length scales.
The perturbation theory is constructed as follows. The fields are expanded 
in series of a small parameter $\epsilon$: $\eta_a=\sum_i\epsilon^i\eta_a^{(i)}$ 
and collected order by order in the functional. The zero-th order is the original 
functional, while the first order is identically zero provided the leading order 
in the series expansion satisfies the equations of motion. Physically relevant 
correction thus appear at the order $\epsilon^2$ of the expanded Ginzburg-Landau 
functional.
The length-scale analysis is done by applying the previously discussed 
perturbative theory to the case where the leading order is the ground-state.
As the ground state is homogeneous, the perturbation operator will 
drastically simplify.
We choose the following perturbative expansion around the ground state 
\Equation{Eq:expansion}{
\eta_a=u_a+\frac{\epsilon f_a}{\sqrt{\kappa_a}}\,,~~
\varphi_{12}=\bvarphi+\epsilon\sqrt{\frac{\kappa_1u_1^2+\kappa_2u_2^2}
{\kappa_1\kappa_2u_1^2u_2^2}}\phi\,.
}
where $u_a$ and $\bvarphi$ denote the ground state and $f_a$, $\phi$, 
the perturbations. Collecting the perturbations in $\Upsilon=(f_1,f_2,\phi)^T$
the term which is second order in $\epsilon$ in the Ginzburg-Landau functional 
reads as:
\Equation{Eq:perturbation:operator}{
\frac{1}{2}\Upsilon^T\left(\Grad^2+{\cal M}^2  \right)\Upsilon\,,
}
where the entries of the (squared) mass matrix are: 
\begin{widetext}
\SubAlign{Eq:perturbation}{
{\cal M}^2_{f_1f_1}&=\frac{2}{\kappa_1}\left(
	\alpha+3\beta u_1^2+(\gamma+\delta\cos2\bvarphi) u_2^2
	+3\omega u_1u_2\cos\bvarphi \right)\\
{\cal M}^2_{f_2f_2}&=\frac{2}{\kappa_2}\left(
	\alpha+3\beta u_2^2+(\gamma+\delta\cos2\bvarphi) u_1^2
	+3\omega u_1u_2\cos\bvarphi\right) \\
{\cal M}^2_{f_1f_2}&=\frac{1}{\sqrt{\kappa_1\kappa_2}}
	\left(4(\gamma+\delta\cos2\bvarphi)u_1u_2
	+(\nu+3\omega(u_1^2+u_2^2))\cos\bvarphi \right)\\
{\cal M}^2_{\phi\phi}&=
\frac{\kappa_1u_1^2+\kappa_2u_2^2}{\kappa_1\kappa_2u_1^2u_2^2}
\left( -4\delta u_1^2u_2^2\cos2\bvarphi		
	-(\nu+\omega(u_1^2+u_2^2))u_1u_2\cos\bvarphi	\right)	\\
{\cal M}^2_{f_1\phi}&=
\sqrt{\frac{\kappa_1u_1^2+\kappa_2u_2^2}{\kappa_1^2\kappa_2u_1^2u_2^2}}
\left( -4\delta u_1u_2^2\sin2\bvarphi		
	-(\nu+\omega(3u_1^2+u_2^2))u_2\sin\bvarphi	\right)	\\
{\cal M}^2_{f_2\phi}&=
\sqrt{\frac{\kappa_1u_1^2+\kappa_2u_2^2}{\kappa_1\kappa_2^2u_1^2u_2^2}}
\left(-4\delta u_1^2u_2\sin2\bvarphi		
	-(\nu+\omega(u_1^2+3u_2^2))u_1\sin\bvarphi \right)	\,.
}
\end{widetext}%
Finally, the length scales are given by finding the eigenstates of 
\Eqref{Eq:perturbation:operator}. More precisely, the eigenvalues 
$m_i^2$ of the (symmetric) mass matrix ${\cal M}^2$, whose elements 
are given in \Eqref{Eq:perturbation}, are the (squared) masses of 
the elementary excitations. The corresponding length scales are 
the inverse (eigen)masses: $\ell_i=1/\sqrt{m_i^2}$.

The length scales that correspond to the phase diagram \Figref{Fig:GroundStateEta} 
are displayed in \Figref{Fig:GroundLength} (only the two largest). 
The largest length scale $\ell_1$ diverges both at $T_c$ and at $T_{\groupZ{2}}$, 
the temperature of the time-reversal symmetry breaking. This is the standard 
mean-field divergence of the coherence length at a two different second order 
transitions. Notice that the second largest length-scale, $\ell_2$ diverges at a 
single point at $T_c$ and where $\lambda=\eta$. This indicates a point of higher 
symmetry in the phase diagram.

\begin{figure*}[!htb]
\hbox to \linewidth{\hss
\includegraphics[width=0.7\linewidth]{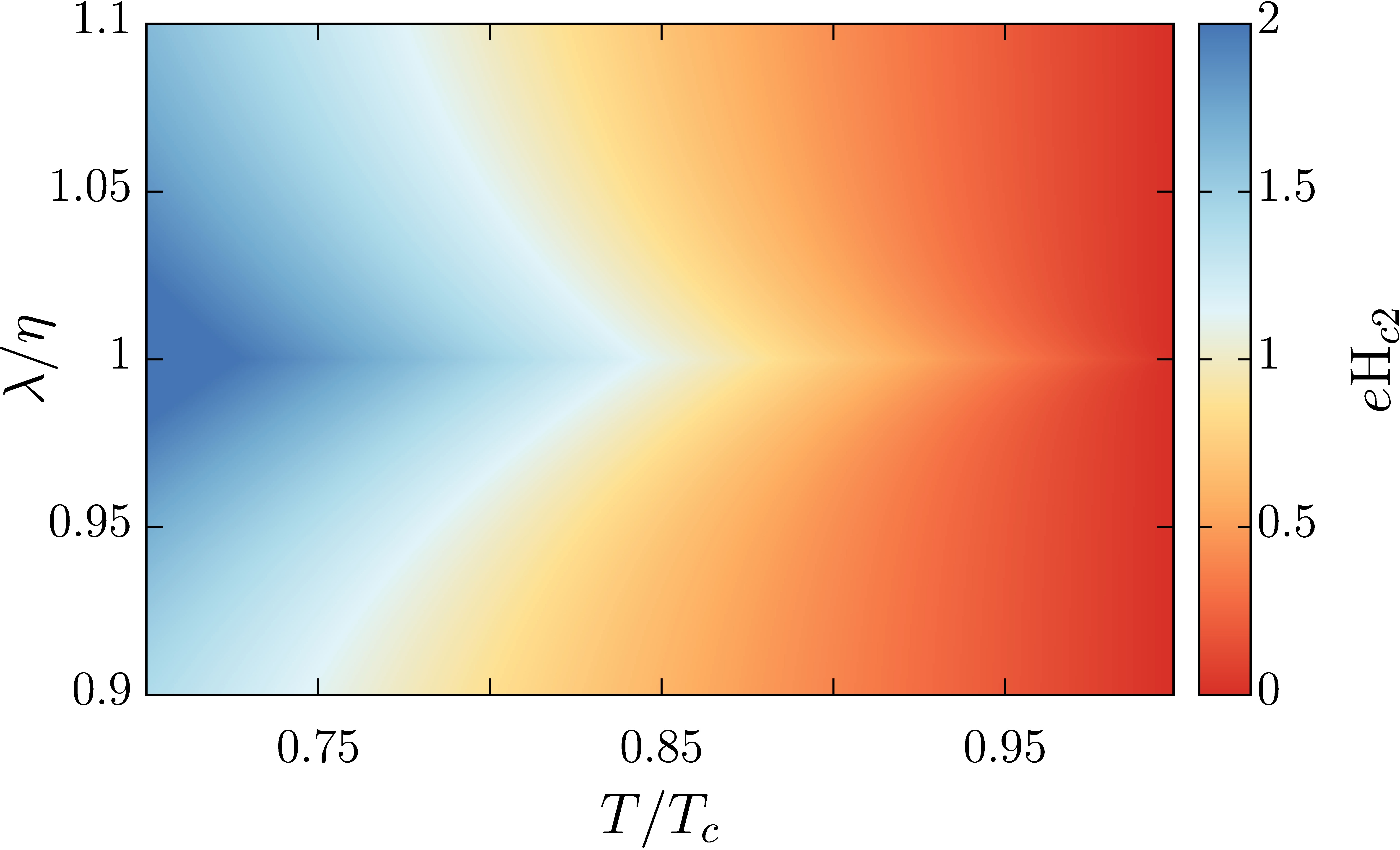}
\hss}
\caption{ (Color online) --
Second critical field calculated from the analysis of the Ginzburg-Landau 
functional \Eqref{Eq:FreeEnergy} with the coefficients determined consistently 
from the microscopic theory.
} \label{Fig:Hc2}
\end{figure*}

Notice that the other relevant length scale of the theory is the (London) penetration 
depth $\lambda_L$ of the magnetic field. It is relatively easy to see that the 
perturbations of the vector potential completely decouple (at the linear level) 
from those of the condensates. The penetration depth is given as the inverse mass of 
$\A$ and simply reads as 
\Equation{}{
\lambda_L=\frac{1}{e\varrho}=
\frac{1}{e\sqrt{\kappa_1|\eta_1|^2+\kappa_2|\eta_2|^2}}\,,
}
which can easily be read of \Figref{Fig:GroundStateEta}. Quite naturally, 
$\lambda_L$ is always finite for $T<T_c$ and diverges when approaching $T_c$. 
So the largest length scale associated with the condensates diverge both at 
$T_c$ and at $T_{\groupZ{2}}$.
It is interesting to note that this automatically imply that $\lambda_L$ is an 
intermediate length scale near $T_{\groupZ{2}}$. This implies that long-range 
inter vortex forces are attractive. This is a necessary, though not sufficient 
condition for non-monotonic interaction between vortices \cite{Carlstrom.Babaev.ea:11,Johan}.

\subsection{Second critical field}
 
The perturbation operator \Eqref{Eq:perturbation:operator} can be used not only 
to determine the relevant length scales of the Ginzburg-Landau theory, but also 
to obtain the second critical field $\Hc{2}$. 
That is by considering the perturbation 
operator around the normal state. More precisely, in the original parametrization 
the normal state is $|\psi_1|=|\psi_2|=0$. Using \Eqref{Eq:Transform}, this implies 
that the normal state in the new variables is $|\eta_1|=|\eta_2|=0$ and thus 
$u_1=u_2=0$ and $\bvarphi=0$.

Close to the second critical field $\Hc{2}$ the magnetic field is approximately 
constant: $\B=B_0{\bs e}_z$ and the densities are small. Thus the Ginzburg-Landau 
equations \Eqref{Eq:EOM:transf} can be linearized around the normal state as 
\Equation{Eq:linearized:hc2:1}{
\Pi^2\Upsilon={\cal M}^2\big\rvert_{u_1=u_2=\bvarphi=0}\Upsilon\equiv{\cal M}_0^2\Upsilon\,.
}
In the Landau Gauge, the vector potential reads as $\A=(0,B_0x,0)^{-1}$. 
As a result, the equations read as 
\Equation{Eq:linearized:hc2:2}{
\big(\Grad^2-(eB_0x)^2\big)\Upsilon={\cal M}_0^2\Upsilon\,.
}
We consider the simple Gaussian ansatz $\Upsilon=C\exp\left(-\frac{x^2}{2\xi^2}\right)$
with the vector $C=(C_1,C_2,0)^T$ and $eB_0=1/\xi^2$. 
 The 
equation \Eqref{Eq:linearized:hc2:2} further simplifies:
\Equation{Eq:linearized:hc2:3}{
{\cal M}_0^2\Upsilon=\frac{-1}{\xi^2}\Upsilon\,.
}
Thus $1/\xi^2$ is an eigenvalue of $-{\cal M}_0^2$. More precisely, its largest: 
\Equation{Eq:linearized:hc2:4}{
e\Hc{2}=\frac{1}{\xi^2}:=
\mathrm{max}\Big(\mathrm{Eigenvalue}\big[-{\cal M}_0^2\big] \Big)\,.
}

It is easy to realize from \Eqref{Eq:perturbation} that the perturbations of the 
relative phase $\Upsilon$ decouple from density perturbations.  The (reduced) mass 
matrix thus becomes:
\Equation{Eq:perturbation:normal}{
{\cal M}_0^2=\left(\begin{array}{cc}
2\alpha/\kappa_1 			& \nu/\sqrt{\kappa_1\kappa_2}			\\ 
\nu/\sqrt{\kappa_1\kappa_2}	& 2\alpha/\kappa_2 	
\end{array}\right)\,,
}
ant its eigenvalues are 
\Equation{Eq:perturbation:normal:EV}{
\frac{\alpha(\kappa_1+\kappa_2)
\pm\sqrt{\alpha^2(\kappa_1-\kappa_2)^2-\nu^2\kappa_1\kappa_2} }{\kappa_1\kappa_2}\,.
}
As a result, we find the second critical field in the dimensionless units of 
Eq.\Eqref{Eq:FreeEnergy}
\Equation{Eq:hc2}{
\Hc{2}=\frac{-\alpha(\kappa_1+\kappa_2)
+\sqrt{\alpha^2(\kappa_1-\kappa_2)^2-\nu^2\kappa_1\kappa_2} }{e\kappa_1\kappa_2}\,.
}

Notice that for example that for a given value of $\kappa_1$, then having $\kappa_2\ll1$ 
implies that the second critical field can become very large. Tracing this back to the 
original parametrization implies that if the prefactor of the mixed gradients $k_{12}$ 
is close to the critical value $\sqrt{k_1k_2}$ where the energy would become unbounded, 
then the second critical field can become arbitrarily large. This limit for example can 
be realized for the microscopic coefficients satisfying $K_1\gg K_2,K_3$. Note that the 
instability never happens since $\kappa_2$ is always positive. 
\Figref{Fig:Hc2} displays the second critical field as a function of the 
parameters of the microscopic model.

\section{Topological excitations} 
\label{Ref:Topology}

The topological properties, as well as the structure of the ground state 
hints to a rich spectrum of topological excitations, or topological defects, 
that may occur in the theory of interest. Indeed, the theory features 
domain-walls that separate between different broken time-reversal phases; 
vortices that can either carry fractional or integer number of flux quanta. 
Moreover vortices can further be characterized by $\groupCP{1}$ invariants 
and in that case they are referred to as skyrmions. All these topological 
excitations are potential observable signatures of the $s+is$ superconducting 
state.

\subsection{Domain-walls}

\begin{figure}[!htb]
\hbox to \linewidth{\hss
\includegraphics[width=.9\linewidth]{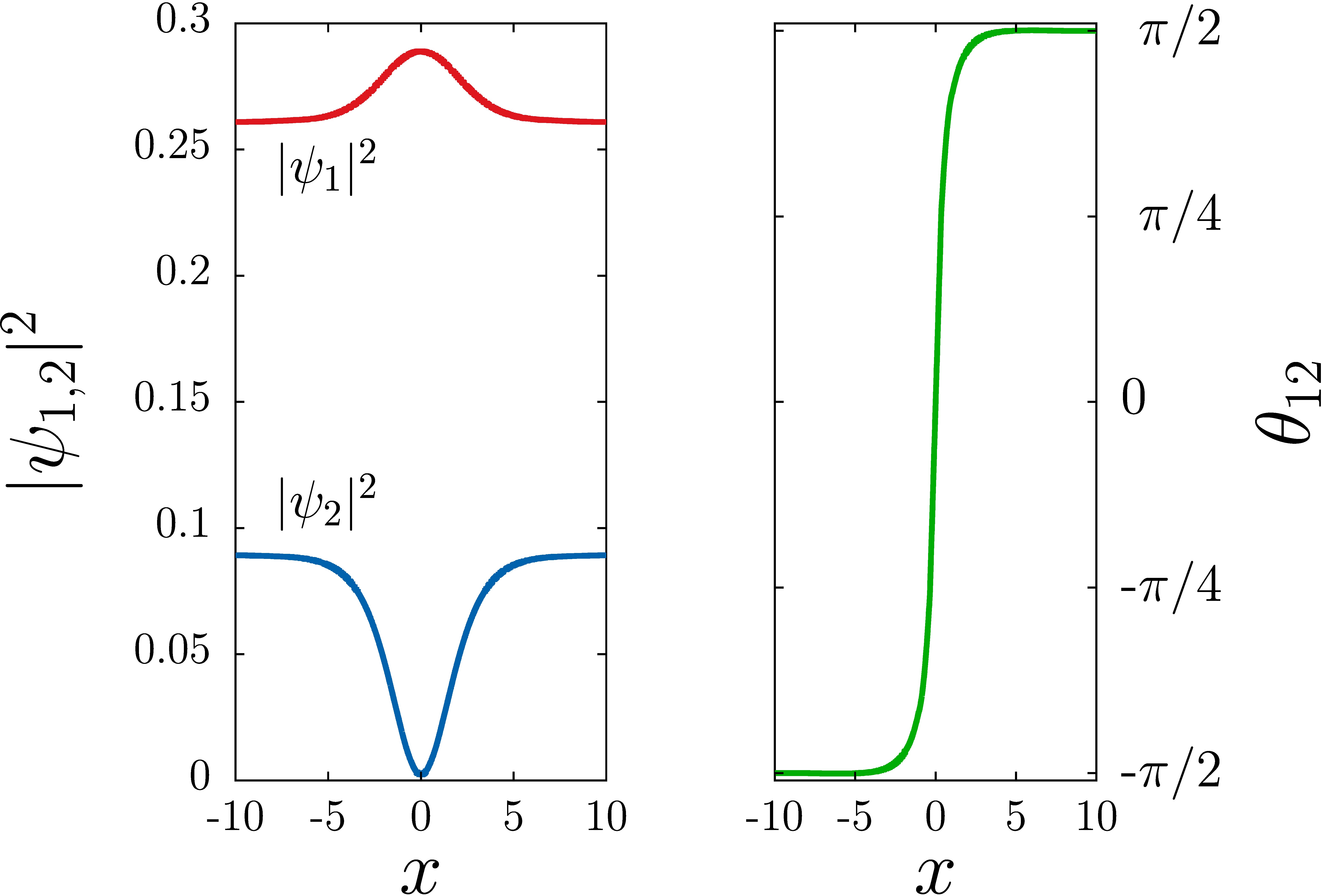}
\hss}
\caption{
(Color online) --
Example of a domain wall that interpolates between two different time-reversal 
symmetry broken states (i.e. $\theta_{12}=\pm\pi/2$). 
}
\label{Fig:DW}
\end{figure}

The ground state phase which is of principal interest here, is the $s+is$ 
superconducting state \Eqref{Eq:GSsis}. It is characterized, besides usual 
spontaneous breakdown of the $\groupU{1}$ gauge symmetry, by a discrete degeneracy 
due to the relative phase being $\theta_{12}=\pm\pi/2$. This discrete degeneracy 
corresponds to the spontaneous of breakdown of the time-reversal symmetry. 
The spontaneous breakdown of a discrete (here $\groupZ{2}$) symmetry is in 
general associated with topological excitations in the form of domain walls that 
interpolate between the two energetically degenerate states. \Figref{Fig:DW} shows 
such a domain wall solution interpolating between $\theta_{12}=-\pi/2$ and 
$\theta_{12}=+\pi/2$.

\subsection{Flux quantization and vortices}

Defining the flux quantization in the original Ginzburg-Landau model 
\Eqref{Eq:FreeEnergy} requires some careful algebraic calculations because of 
the mixed gradient terms. This operation becomes elementary in the rewritten 
Ginzburg-Landau model \Eqref{Eq:FreeEnergy:transf}. Indeed the magnetic flux, 
for example through the plane, reads as:
\SubAlign{Eq:Quantization}{
\Phi&=\int_{\mathbb{R}^2}\B\cdot d{\bf S}=\oint_{\cal C} \A\cdot d{\bs\ell}  
\label{Eq:Quantization:a}\\
&=\oint_{\cal C}\frac{\J\cdot d{\bs\ell}}{e^2\varrho^2}-
\sum_a\frac{\kappa_a|\eta_a|^2}{e\varrho^2}\Grad\varphi_a\cdot d{\bs\ell}
\label{Eq:Quantization:b} \\
&=-\oint_{\cal C}\sum_a\frac{\kappa_a|\eta_a|^2}{e\varrho^2}
\Grad\varphi_a\cdot d{\bs\ell} 
\label{Eq:Quantization:c} 
}
where $\cal C$ is the contour of the $\mathbb{R}^2$ at spatial infinity.
To obtain \Eqref{Eq:Quantization:b}, we use the relation \Eqref{Eq:Currents2} 
to express the vector potential in terms of the current and phase gradients. 
Because of the spontaneous breakdown of the $\groupU{1}$ symmetry, the vector 
potential is massive and so is the current (Meissner effect). The current 
should vanish asymptotically (for the energy to be finite) and thus the 
flux reads as in \Eqref{Eq:Quantization:c}.
The phase of the complex fields $\varphi_a$ can wind only an integer number 
of times $n_a$ and thus $\oint_{\cal C}\Grad\varphi_a=2\pi n_a$. It follows 
that if both components have the same winding number $n_1=n_2=n$, then the 
flux is quantized: $\Phi=n\Phi_0$ with the flux quantum $\Phi_0=-2\pi/e$.

There is no formal reason why both $\eta_{1,2}$ should have the same winding 
number $n_{1,2}$. In that case, the resulting configuration carries only a 
fraction of the flux quantum. For example consider the simplest case where 
$n_1=1$ and $n_2=0$. Then the configuration carries only a fraction 
$\kappa_1|\eta_1|^2/\varrho^2$ of the flux quantum. Hence it is called a 
fractional vortex. However, because this gives a nonzero winding of the 
relative phase in \Eqref{Eq:GLRewritten} this implies that such a configuration 
have a logarithmically divergent energy. Indeed, the part of the free energy 
containing the gradients of the relative phase is 
$\int\frac{\kappa_1\kappa_2|\eta_1|^2|\eta_2|^2}{2\varrho^2}(\Grad\varphi_{12})^2 $. 
At long range, i.e. at a given distance $r$ for the core, densities are approximately 
constant and thus its contribution is approximated by 
\Equation{Eq:Fractional}{
\sim \int_{r_0}^r r^\prime dr^\prime (\frac{1}{r^\prime}\partial_\theta\varphi_{12})^2
\sim \int_{r_0}^r \frac{dr^\prime}{r^\prime}
\sim\log \frac{r}{r_0}	\,.
}
This divergence, which puts an heavy energy penalty on fractional vortices, 
is absent if both components have the same winding. As a result, configurations 
with fractional flux cannot be excited in bulk superconductors. Note however 
they can be stabilized near boundaries \cite{Silaev:11}, in mesoscopic 
samples \cite{Babaev:02,Bluhm.Koshnick.ea:06,Chibotaru.Dao.ea:07,Chibotaru.Dao:10,
Geurts.Milosevic.ea:10} 
or in samples with geometrically trapped domain walls \cite{Garaud.Babaev:14}.
Note that the condition for flux quantization still allows non-trivial 
configurations of the magnetic field.

\subsection{Additional \texorpdfstring{$\groupCP{1}$}{CP1} invariants -- Skyrmions}

As discussed in the next subsection, the topological properties of the model 
can also be understood using the mapping to a nonlinear $\sigma$-model. 
In contrast to the topological invariant characterizing vortices (i.e. the 
winding number which is defined as a line integral over a closed path), an 
additional topological $\groupCP{1}$ index which is in general associated 
with skyrmionic excitations can be defined as an integral over the plane 
\cite{Garaud.Carlstrom.ea:13}. Defining the complex vector ${\bs\eta}$ as 
${\bs\eta}^\dagger=(\sqrt{k_1}\eta_1^*,\sqrt{k_2}\eta_2^*)$, and provided 
${\bs\eta}\neq0$, the topological $\groupCP{1}$ index is given as an integral 
over the $xy$-plane:
\Equation{Eq:Charge2}{
   \Q({\bs\eta})=\int_{\mathbb{R}^2}\frac{i\epsilon_{ji}}{2\pi|{\bs\eta}|^4} \Big[
   |{\bs\eta}|^2\partial_i{\bs\eta}^\dagger\partial_j{\bs\eta}
   +{\bs\eta}^\dagger\partial_i{\bs\eta}\partial_j{\bs\eta}^\dagger{\bs\eta}
   \Big]dxdy\,.
}
Provided ${\bs\eta}\neq0$, the $\groupCP{1}$ index $\Q$ is an integer number 
and is equal to the number of flux quanta: $\Q=\int B/\Phi_0=n$ ($\Phi_0$ being 
the flux quantum and $n$ the number of flux quanta) \cite{Garaud.Carlstrom.ea:13}. 
As a result, in the case of an axially symmetry vortex with a core where all 
superconducting condensates simultaneously vanish (i.e. ${\bs\eta}=0$), then 
$\Q=0$. On the other hand, if singularities happen at different locations 
(i.e. ${\bs\eta}\neq0$), then $\Q\neq0$ and the quantization condition holds. 
As a result, $\Q$ is a useful quantity that can discriminate between vortices 
and skyrmions (which are coreless defects). See Ref.~\cite{Garaud.Carlstrom.ea:13} 
for a rigorous discussion .
One should note that the flux-quantization condition \Eqref{Eq:Quantization} 
and the integral formula for the topological charge $\Q$ above are valid 
only for field configurations for which ${\bs\eta}$ never vanishes. Note that 
flux is also quantized for ordinary vortices, for which ${\bs\eta}$ vanishes, 
but then it is no longer associated with the topological charge $\Q$, 
but with a $U(1)$ topological invariant (the usual winding number).

\subsection{Mapping to a nonlinear \texorpdfstring{$\sigma$}{sigma}-model} 

The mapping to a nonlinear $\sigma$-model consists of rewriting the theory in 
term a massive $\groupU{1}$ vector field (the current) coupled to a compact 
$\groupO{3}$ vector. Importantly, in this kind of mapping the supercurrent is
coupled to Faddeev-Skyrme terms \cite{Babaev.Faddeev.ea:02}. Starting from the 
theory in terms of charged and neutral modes \Eqref{Eq:GLRewritten}, it is possible 
to map the two-component model to an easy-plane non-linear $\sigma$-model. 
This mapping is done by defining the pseudo-spin unit vector $\bf n$ as a projection 
of the superconducting degrees of freedom onto spin-$1/2$ Pauli matrices $\bs\sigma$: 
\Equation{Eq:Projection}{
 {\bf n} =\frac{{\bs\eta}^\dagger\bs \sigma{\bs\eta}}{{\bs\eta}^\dagger{\bs\eta}}
 \,,  ~~\text{where}~~
 {\bs\eta}^\dagger=(\sqrt{k_1}\eta_1^*,\sqrt{k_2}\eta_2^*)\, .
}
The following identity is useful to rewrite the free energy \Eqref{Eq:GLRewritten} 
in terms of the pseudo-spin $\bf n$, total density $\varrho^2:={\bs\eta}^\dagger{\bs\eta}$ 
and gauge invariant current $\J$
\Align{Eq:NLS-Identity1}{
\frac{\varrho^2}{4}\partial_kn_a\partial_kn_a+(\Grad\varrho)^2=& 
\frac{\kappa_1\kappa_2|\eta_1|^2|\eta_2|^2}{\varrho^2}(\Grad\varphi_{12})^2 
\nonumber \\
&+\sum_a\kappa_a(\Grad|\eta_a|)^2 \,,
}
where summation on repeated indices is implied. Using the definition of the current 
and noting that 
\Equation{Eq:NLS-Identity2}{
4\varepsilon_{ijk}\partial_i\left(
\sum_a\frac{\kappa_a|\eta_a|^2}{\varrho^2}\partial_j\varphi_a\right) = 
\varepsilon_{ijk}\varepsilon_{abc}n_a\partial_in_b\partial_jn_c ,
}
where $\varepsilon$ is the Levi-Civita symbol, the magnetic field reads as
\Equation{Eq:NLS-Identity3}{
B_k=\frac{1}{e}\varepsilon_{ijk}\left(\partial_i
\left( \frac{J_j}{e\varrho^2} \right)
-\frac{1}{4}\varepsilon_{abc}n_a\partial_in_b\partial_jn_c \right)\,,
}
and the free energy \Eqref{Eq:GLRewritten} can be written as
\Align{Eq:NLSM:sis}{
 \F&= \frac{1}{2}(\Grad\varrho)^2
 +\frac{\varrho^2}{8}\partial_k n_a\partial_k n_a
 +\frac{\J^2}{2e^2\varrho^2}+V(\varrho,{\bs n})	\nonumber \\
+&\frac{1}{2e^2}\left[\varepsilon_{ijk}\left(
\partial_i \left(\frac{J_j}{e\varrho^2}\right)
	-\frac{1}{4}\varepsilon_{abc}n_a\partial_i n_b\partial_j n_c \right)\right]^2
 \,,
}
where $V(\varrho,{\bs n})$ is the potential terms 
\Equation{Eq:NLSM:potential}{
V=\frac{\varrho^2}{2}(a_1+a_2n_x)
+\frac{\varrho^4}{4}(b_1+2b_2n_x+b_3n_x^2+b_4n_z^2)\,,
}
with the coefficients
\Align{Eq:NLSM:potentialCoeff}{
b_1&=\beta+\gamma-4\delta\,,~
b_2=2\omega\,,~
b_3=8\delta\,,~
b_4=\beta-\gamma+4\delta\,,\nonumber \\
a_1&=2\alpha\,,~~
a_2=2\nu\,.
}
The pseudo-spin is a map from the one-point compactification 
of the plane ($\mathbb{R}^2\simeq S^2 $) to the two-sphere target space 
spanned by $\bf n$. That is ${\bf n}: S^2\to S^2$, classified by 
the homotopy class $\pi_2(S^2)\in\mathbb{Z}$, thus defining the integer 
valued topological (skyrmionic) charge  
\Equation{Eq:Charge}{
   \Q({\bf n})=\frac{1}{4\pi} \int_{\mathbb{R}^2}
   {\bf n}\cdot\partial_x {\bf n}\times \partial_y {\bf n}\,\,
  dxdy \,.
}
Heuristically, the topological charge \Eqref{Eq:Charge} can be understood 
as the integer that counts the number of times the pseudo-spin wraps the 
target sphere. If a field configuration spans only a portion of the target 
sphere, then the associated flux needs not to be quantized. 
Note that this definition of the topological charge \Eqref{Eq:Charge}, 
is actually equivalent to that given earlier \Eqref{Eq:Charge2}.

\section{Conclusion}

To summarize, we presented a microscopic derivation for $N$-component 
Ginzburg-Landau models with a focus on the time-reversal symmetry breaking 
$s+is$ state for a three-band microscopic model. This model is widely believed 
to describe hole-doped 122 compounds. We consistently derived the two-component 
Ginzburg-Landau functional that is relevant for the case of an interband-dominated 
pairing.
We discussed the elementary properties of these models: normal modes and length scales, 
critical fields and basic topological defects.

\begin{acknowledgments}
The work was supported by the Swedish Research Council Grants
No. 642-2013-7837.
The computations were performed on resources provided by the
Swedish National Infrastructure for Computing (SNIC) at National
Supercomputer Center at Link\"oping, Sweden.
\end{acknowledgments}
\section*{References}
%

\end{document}